
\font\rm=cmr12
\font\bf=cmbx12

\rm
\baselineskip 24pt
\hsize 6.5in
\vsize 8.5in
\def\lppm{\Lambda\rightarrow p\pi^-}
\def\lnpo{\Lambda\rightarrow n\pi^o}
\def\loppm{\Lambda^o\rightarrow p\pi^-}
\def\lonpt{\Lambda^o\rightarrow n\pi_3}
\def\lonpe{\Lambda^o\rightarrow n\pi_8}
\def\spppo{\Sigma^+\rightarrow p\pi^o}
\def\spnpp{\Sigma^+\rightarrow n\pi^+}
\def\smnpm{\Sigma^-\rightarrow n\pi^-}
\def\spppt{\Sigma^+\rightarrow p\pi_3}
\def\spppe{\Sigma^+\rightarrow p\pi_8}
\def\soppm{\Sigma^o\rightarrow p\pi^-}
\def\sonpt{\Sigma^o\rightarrow n\pi_3}
\def\xlpo{\Xi^o\rightarrow \Lambda\pi^o}
\def\xlpm{\Xi^-\rightarrow\Lambda\pi^-}
\def\xlopm{\Xi^-\rightarrow\Lambda^o\pi^-}
\def\xlopt{\Xi^o\rightarrow\Lambda^o\pi_3}
\def\xlope{\Xi^o\rightarrow\Lambda^o\pi_8}
\def\xsopt{\Xi^o\rightarrow\Sigma^o\pi_3}
\def\xsopm{\Xi^-\rightarrow\Sigma^o\pi^-}
\def\oxmpo{\Omega\rightarrow\Xi^-\pi^o}
\def\oxopm{\Omega\rightarrow\Xi^o\pi^-}
\def\oxmpt{\Omega\rightarrow\Xi^-\pi_3}
\def\oxmpe{\Omega\rightarrow\Xi^o\pi_8}
\def\newref{\vskip .02in\noindent}
\def\fpi{f_\pi}
\def\tmnpo{(2m_N+\omega )}
\def\mnpml{(m_N+m_\Lambda )}
\def\mnpms{(m_N+m_\Sigma )}
\def\mlpmx{(m_\Lambda +m_\Xi )}
\def\mspmx{(m_\Sigma +m_\Xi )}
\def\mlmmn{(m_\Lambda -m_N)}
\def\msmmn{(m_\Sigma -m_N)}
\def\mxmml{(m_\Xi -m_\Lambda )}
\def\mxmms{(m_\Xi -m_\Sigma )}
\def\dioh{$\Delta I={1\over 2}\ $}
\def\dith{$\Delta I={3\over 2}\ $}
\def\cloh{c_\Lambda (1/2)}
\def\clth{c_\Lambda (3/2)}
\def\dsthoh{D_\Sigma^{3/2} (1/2)}
\def\dsohoh{D_\Sigma^{1/2} (1/2)}
\def\sqot{\sqrt{1\over 3}}
\def\sqtt{\sqrt{2\over 3}}
\def\sqth{\sqrt{3}}
\def\sqs{\sqrt{6}}
\def\sqt{\sqrt{2}}
\def\mksmmps{(m_K^2-m_\pi^2)}
\def\omdm{(\omega -\delta m)}
\def\opdm{(\omega +\delta m)}
\def\dpp{d^{\prime\prime}}
\def\fpp{f^{\prime\prime}}
\def\lamo{\Lambda^o}
\def\sigo{\Sigma^o}
\def\sigtxt{$\Sigma\ $}
\def\pit{\pi_3}
\def\pie{\pi_8}
\def\thetam{\theta_m}
\def\lamtxt{$\Lambda\ $}
\def\pio{\pi^o}

\
\vskip .5in
\centerline{\bf Strong Isospin Mixing Effects on the Extraction of \dith
Non-Leptonic}
\centerline{\bf Hyperon Decay Amplitudes}

{\hfill ADP-95-4/T171}
\vskip .5in
\centerline{Kim Maltman}
\vskip .15in
\centerline{Department of Mathematics and Statistics, York University,
4700 Keele St.,}
\centerline{North York, Ontario, CANADA  M3J 1P3}
\centerline{and}
\centerline{Department of Physics and Mathematical Physics, University of
Adelaide}
\centerline{Adelaide, South Australia 5005, Australia}
\vskip .5in
\centerline{\bf ABSTRACT}
\vskip .15in
\noindent
The existence of isospin admixtures in the physical $\Lambda$, $\pi^o$
complicates the extraction of \dith non-leptonic hyperon decay amplitudes
from experimental data, allowing contributions associated with large
\dioh amplitudes to appear in the (nominally) \dith amplitudes
obtained ignoring these admixtures.  We show how to correct for this
effect to leading order in $(m_d-m_u)$ and extract the true \dith
amplitudes.  The resulting corrections are modest ($<25\%$) for
s-waves, but extremely large, $\simeq 100\%$ and $\simeq 400\%$
for $\Lambda$ and $\Xi$ p-waves, respectively.
\vfill\eject
In a recent paper$^1$, Karl has demonstrated that the existence of
$\lamo -\sigo$ mixing produces small but non-trivial corrections to
$g_V$, $g_A$ in hyperon semi-leptonic decays.  Here we investigate a
analogous effects which alters the extraction of \dith non-leptonic hyperon
decay amplitudes from experimental data.

Recall that the physical \lamtxt and $\pi^o$ are admixtures of the
pure $I=0,1$ isospin states $\Lambda^o$, $\Sigma^o$ and $\pi_8$, $\pi_3$,
respectively (we use throughout the notation $\Lambda$, $\pi^o$,
$\eta$ for the physical, mixed-isospin states, and $\Lambda^o$,
$\Sigma^o$, $\pit$, $\pie$ for the pure isospin states).  Since the
isospin impurities are small, $O(10^{-2})$, we may write
$$\eqalignno{&\Lambda =\lamo +\theta_b\sigo\cr
&\pio =\pit +\thetam\pie\ .&(1)\cr}$$
To leading order in the current quark masses, adopting the phase
conventions of Ref.~2, one has $^{3,4}$
$$\thetam =-\theta_b ={\sqth\over 4}\Biggl[ {(m_d-m_u)\over
[m_s-(m_u+m_d)/2]}\Biggr]\ .\eqno(2)$$
Following Ref.~1, all numbers quoted below will be based on the value
$\thetam \simeq 0.015$.

It is immediately obvious that the isospin admixtures in Eqn.~(1)
invalidate the usual procedure for extracting \dith contributions
to amplitudes from experimental data.  To illustrate, consider the
case of \lamtxt decay.  The physical (s- or p-wave) amplitudes
are given, in terms of the corresponding amplitudes involving only
pure isospin states (which we shall henceforth call ``isospin-purified''
amplitudes) by
$$\eqalignno{&M(\lppm )=M(\loppm )-\thetam M(\soppm )\cr
&M(\lnpo )=M(\lonpt )+\thetam \bigl[ M(\lonpe )-M(\sonpt )\bigr]\ .&(3)\cr}$$
We know empirically that hyperon decay amplitudes are dominantly \dioh and can
therefore ignore the small \dith components of the \sigtxt amplitudes
in the correction terms in Eqn.~(3).  Let us write $\Delta I={1\over 2}$,
${3\over 2}$ decompositions for the $\lamo$ amplitudes
$$\eqalignno{&M(\loppm )=\sqtt\cloh -\sqot\clth\cr
&M(\lonpt )=-\sqot\cloh -\sqtt\clth\ ,&(4)\cr}$$
where $\cloh$, $\clth$ are the \dioh and \dith reduced matrix elements
(with $c_\Lambda\rightarrow A_\Lambda$, $B_\Lambda $ for s- and
p-waves, respectively), and use the \dioh relations
$$\eqalignno{&M(\soppm )=-M(\spppt )={\sqt\over 3}\dsthoh -{\sqt\over 3}
\dsohoh\cr
&M(\sonpt )={1\over 2}\bigl[ M(\spnpp )+M(\smnpm )\bigr] =
{2\over 3}\dsthoh +{1\over 3}\dsohoh &(5)\cr}$$
for the \sigtxt amplitudes, where $\dsthoh$ and $\dsohoh$ are the
reduced matrix elements of the \dioh transition operator for final
$N\pi$ isospins ${3\over 2}$ and ${1\over 2}$, respectively, and
we have ignored the small \dith parts of the transitions.  The \dith
component of the isospin-purified amplitudes can be extracted by
forming the combination
$$\clth =-\sqot M(\loppm )-\sqtt M(\lonpt )\ .\eqno(6)$$
If, however, we form the analogous combination of the physical amplitudes,
we obtain
$$-\sqot M(\lppm )-\sqtt M(\lnpo )=\clth -\thetam \Bigl[ \sqtt M(\lonpe )
-\sqtt \dsthoh\Bigr]\ .\eqno(7)$$
The terms proportional to $\thetam$ in Eqn.~(7), though pure \dioh ,
enter the nominal \dith combination.  Since $\thetam$ is of the same
order as the normally extracted \dith to \dioh amplitude ratio$^{2,5}$,
these corrections may, in general, be expected to produce significant
errors in extracting the true \dith amplitudes. Similar corrections
are present in the \dith $\Xi$ relation and the
$\Delta I={1\over 2}$-rule-violating \sigtxt triangle relation.

To obtain the corrections necessary to convert the physical
to the isospin-purified amplitudes, and hence to obtain the true
\dith contributions, we require the amplitudes for the processes
$\sonpt$, $\soppm$, $\lonpe$, $\spppe$, $\xlope$, $\xsopm$, $\xsopt$ and
$\oxmpe$.  The first two of these can be obtained, to good accuracy,
from the physical amplitudes for $\spnpp$, $\spppo$ and $\smnpm$,
using the \dioh relations of Eqn.~(5).  The remaining
correction amplitudes, however, are not observable, and must be obtained
theoretically.  We treat the s- and p-waves separately.

For the s-waves, it is well-known that a lowest order chiral SU(3) analysis
provides an excellent fit to the experimental amplitudes$^{2,6}$.  We,
therefore, take the desired correction amplitudes from the same analysis.
In terms of the usual $F$, $D$ parameters one obtains
$$\eqalignno{&A(\lonpe )=-\sqth (3F+D)/2\fpi\cr
&A(\sonpt )=-\sqth (F-D)/2\fpi\cr
&A(\soppm )=-\sqs (F-D)/2\fpi\cr
&A(\spppe )=-3\sqt (F-D)/2\fpi\cr
&A(\xsopm )=\sqs (F+D)/2\fpi\cr
&A(\xsopt )=\sqth (F+D)/2\fpi\cr
&A(\xlope )=\sqth (3F-D)/2\fpi&(8)\cr}$$
with $\fpi\simeq 93$ MeV the $\pi$ decay constant and ${F\over 2\fpi} =
-0.92\times 10^{-7}$,
$D/F=-0.42$.  The resulting corrections are presented in Table 1, where we
display the experimental amplitudes, together with the corrections to be
added to convert them to the corresponding
isospin-purified amplitudes.  Extracting, then,  the true
\dith combinations, $A_\Lambda (3/2)$, $A_\Xi (3/2)$ and
$\Delta_\Sigma^A = A(\spnpp )-A(\smnpm )-\sqt A(\spppt )$
one finds
$$\eqalignno{&A_\Lambda (3/2)=0.059\times 10^{-7}-0.0047\times 10^{-7}
\cr
&A_\Xi (3/2)=-0.227\times 10^{-7}-0.050\times 10^{-7}\cr
&\Delta_\Sigma^A=0.499\times 10^{-7}+0.177\times 10^{-7}&(9)\cr}$$
where, in all cases, the first number is obtained using
the uncorrected physical amplitudes and the second is the correction
resulting from using, instead, the isospin-purified amplitudes.
The corrections, in this case, are modest, no more than $\simeq 25\%$.

The situation for the p-wave amplitudes is rather different.  Here, as is
well-known, the leading contributions are expected to be due to baryon
pole graphs produced through parity-conserving (PC) weak baryon-baryon
transitions.  The SU(3) parametrization of these transitions
obtained from the leading soft-pion analysis of the s-wave
decays, however, fails miserably in accounting for the p-wave amplitudes.
A reasonable fit, including now small $K$ pole contributions, can be
obtained$^{7,2}$ only by using a considerably different SU(3)
parametrization.  Even then the fit is not nearly as good as in the s-wave case
and, although ideas exist for explaining the apparent
discrepancy between s- and p-wave fit parameters$^2$, the theoretical
situation is not at all clear, giving us somewhat less confidence in the
extraction of values for the unobservable amplitudes.
In order to investigate theoretical uncertainties
we will, therefore, also evaluate
the p-wave correction amplitudes using the model
of Ref.~8, which includes $K$ pole as well as ${1\over 2}^+$ and
${1\over 2}^{+*}$ baryon pole contributions to the amplitudes.  This model
actually provides a somewhat better numerical fit to the
data, though at the cost of employing a $K$-to-$\pi$ weak transition
strength an order of magnitude greater than that extracted from
$K\rightarrow\pi\pi$, which makes it appear somewhat suspect.  As we
will see, the resulting corrections to the extracted \dith amplitudes
turn out to be rather similar in the two cases, giving us improved
confidence in our numerical results, despite the increased theoretical
uncertainty.

We begin with the fit of Refs.~2,7.  $F_A$, $D_A$ are the axial vector
coupling parameters ($F_A+D_A=1.25$) relevant to the pseudovector couplings
of the pseudoscalar octet to the baryon octet, and $f$, $d$ those for
the baryon-baryon weak couplings.  The parametrization of Ref.~7 is
$D_A/F_A=1.8$, $d/f=-0.85$, $f=4.7\times 10^{-5}$ MeV.  The expressions
for the ground state baryon pole
contributions to the isospin-purified versions of the observed p-wave
amplitudes are given in Ref.~8 (Eqns.~(3.2) and (3.9)).  To convert these
expressions to our conventions, the $\fpi$ of Ref.~8 must be replaced
by $\sqt\fpi$.  There is also a typographical error in the
second term of the last of Eqns.~(3.9), where $(3f+d)$ should read
$(3f-d)$.  The ground state pole contributions to those correction
amplitudes not quoted in Ref.~8, and not obtainable from the $\Sigma^o$
\dioh relations, are then
$$\eqalignno{&B_p(\lonpe )={\mnpml\over 6\fpi}\Bigl[ {(3f+d)(3F_A+D_A)
\over\mlmmn}\Bigr]\cr
&B_p(\spppe )=\sqrt{3\over 2} {\mnpms\over\fpi}\Bigl[ {(f-d)(F_A-D_A)\over
\msmmn}\Bigr]\cr
&B_p(\xlope )=-{\mlpmx\over 6\fpi}\Bigl[ {(3f-d)(3F_A-D_A)\over\mxmml}
\Bigr]\cr
&B_p(\xsopt )=-{\mspmx\over 6\fpi}\Bigl[ {3(f+d)(F_A-D_A)\over\mxmms}
+{2(3f-d)D_A\over\mxmml}\Bigr]\cr
&B_p(\xsopm )=-{\mspmx\over \sqt\fpi}\Bigl[ {(f+d)(F_A+D_A)\over\mxmms}
\Bigr]\ .&(10)\cr}$$
The $K$ pole contributions can be obtained in terms of $D_A$, $F_A$ and the
$K$-to-$\pi$ transition matrix element $a_{K\pi}=<\pi^-\vert H_{PC}^{weak}
\vert K^->$, if one takes the $K$-to-$\pi$ and $\bar K^o$-to-$\pi_8$
transitions elements to be given by the lowest order chiral effective
Lagrangian$^{2,6}$.  Using $K\rightarrow\pi\pi$
data$^2$, and dropping again the small \dith contributions,
one finds $a_{K\pi}=3.18\times
10^{-3}$ MeV$^2$, and $<\pio\vert H_{PC}^{weak}\vert \bar K^o>=-a_{K\pi}
/\sqt $, $<\pie\vert H_{PC}^{weak}\vert \bar K^o>=-a_{K\pi}/\sqs$.
(A numerically very similar relation between the $\pi_{3,8}$ matrix
elements results from estimating them using the QCD-evolved effective
weak Hamiltonian in the factorization approximation.)  The $K$ pole
contributions to the isospin-purified versions of the observed amplitudes
are then as quoted in the ``Fit b'' column of Table 6.10 of Ref.~2, while
the corresponding contributions to the correction amplitudes are
obtainable, for $\sigo$, from the \dioh relations,  and
otherwise, from
$$\eqalignno{&B_K(\lonpe )=-{a_{K\pi}\over 6\sqt}\Bigl[ {\mnpml\over\mksmmps}
\Bigr] {(3F_A+D_A)\over f_K}\cr
&B_K(\spppe )={a_{K\pi}\over 2\sqth}\Bigl[ {\mnpms\over\mksmmps}
\Bigr] {(D_A-F_A)\over f_K}\cr
&B_K(\xlope )={a_{K\pi}\over 6\sqt}\Bigl[ {\mlpmx\over\mksmmps}
\Bigr] {(3F_A-D_A)\over f_K}\cr
&B_K(\xsopt )=-{a_{K\pi}\over 2\sqt}\Bigl[ {\mspmx\over\mksmmps}
\Bigr] {(F_A+D_A)\over f_K}\cr
&B_K(\xsopm )=-{a_{K\pi}\over 2}\Bigl[ {\mspmx\over\mksmmps}
\Bigr] {(F_A+D_A)\over f_K}&(11)\cr}$$
where the overall sign has been adjusted as in
Ref.~2.  The $K$ pole terms are, in all cases, much
smaller than the baryon pole terms.  The resulting total
amplitudes are listed in column 1 of Table 2, the corresponding
physical amplitudes (where such exist) in column 3.  The
$\Sigma^o$ amplitudes are obtained using the \dioh relations, rather
than from the model.

For the alternate
model of Ref.~8, the ground state pole contributions employ the
parametrization $F_A=0.43$, $D_A=0.82$ and $f/d=-1.5$, with
$f=3.9\times 10^{-5}$ MeV (the $f$,$d$ values being obtained from a fit
to the s-wave amplitudes which includes $70^-$ baryon pole terms in
addition to the usual commutator terms), and are given formally by
Eqns~(3.29) of Ref.~8 and Eqns.~(10) above (recall the difference in
conventions for $\fpi$).  The $K$ pole contributions can
be obtained from those above by simply rescaling by the ratio, $-10.7$,
of $a_{K\pi}$ in the two models.
The remaining contributions, associated with the
${1\over 2}^{+*}$ baryon poles, are given in terms of $F^*$, $D^*$
values for the $BB^*\pi$ couplings, $\fpp$, $\dpp$ values for the
$<B^*\vert H_{PC}\vert B>$ couplings, the mean splitting, $\omega\simeq
500$ MeV of the ${1\over 2}^+$ and ${1\over 2}^{+*}$ multiplets, and
the mean splitting, $\delta m\simeq 200$ MeV, of baryons in a given
multiplet differing by one unit of strangeness.  The relations
$\dpp /\fpp =-1$ and $F^*/D^*=1.91$ are assumed, $F^*$ is fit to
the $P_{11}(1440)$ decay width, and $\dpp=-4.4\times 10^{-5}$ MeV
obtained by optimizing the the p-wave amplitude fit.  The contributions
of these poles to the isospin-purified versions of the observed amplitudes
are given by Eqns.~(3.2), (3.21) of Ref.~8 and those to the
correction amplitudes not obtainable using the $\Sigma^o$
\dioh relations by
$$\eqalignno{
&B_*(\lonpe )={2\dpp\mnpml\over 3\sqt\tmnpo}\Bigl[ {(3F^*-D^*)\over\omdm}
   -{2D^*\alpha\over\opdm}\Bigr]\cr
&B_*(\spppe )={2\dpp\mnpms\over \sqth\tmnpo}\Bigl[ {(3F^*-D^*)\over\omdm}
   +{2D^*\alpha\over\opdm}\Bigr]\cr
&B_*(\xlope )={4\dpp\mlpmx\over 3\sqt\tmnpo}\Bigl[ {2D^*\alpha\over\omdm}
   +{(3F^*+D^*)\beta\over\opdm}\Bigr]\cr
&B_*(\xsopt )=-{8\dpp\mspmx\over 3\sqt\tmnpo}\Bigl[ {D^*\alpha\over\omdm}
   \Bigr]\cr
&B_*(\xsopm )=0&(12)\cr}$$
where $\alpha =\tmnpo /(2m_N+\omega +2\delta m)=0.86$ and
$\beta =\tmnpo /(2m_N+\omega +4\delta M)=0.75$.  The total amplitudes
are listed in column 2 of Table 2.
The $\Sigma^o$ amplitudes are again taken, not from the
model, but from the experimental amplitudes, using the \dioh relations.
As mentioned above, the model of Ref.~8 actually appears somewhat
suspect, in view of the large value of $a_{K\pi}$.  Moreover, as
can be seen from Table II of Ref.~8, although the numerical fit to the
p-wave amplitudes is quite reasonable, there is considerable cancellation
amongst the three independent contributions.  The point of employing the
model is simply to test the potential model-dependence of the computed
corrections to the physical amplitudes.

We tabulate, in Table 3, the predicted corrections to the experimental
p-wave amplitudes.  Column 1 re-lists, for convenience, the experimental
values, while columns 2,3 contain the corrections obtained using the
models of Refs.~2,7 (8), respectively.  The results of column 2(3) are
to be added to those of column 1 to convert from the experimental to
the isospin-purified versions of the amplitudes in question.  While
there are non-trivial differences in the predicted corrections for the
$\spppo $ and $\xlpm$ amplitudes in the two models, when we calculate
the p-wave \dith quantities, $B_\Lambda (3/2)$, $B_\Xi (3/2)$
and $\Delta_\Sigma^B$, we find for the two models
$$\eqalignno{&B_\Lambda (3/2)=0.141\times 10^{-7}
+0.526\times 10^{-7}(0.566\times 10^{-7})\cr
&B_\Xi (3/2)=0.530\times 10^{-7}+0.755\times 10^{-7}(0.643\times
10^{-7})\cr
&\Delta_\Sigma^B=5.92\times 10^{-7}-0.77\times 10^{-7}(0.42\times
10^{-7})&(13)\cr}$$
where, as for the s-wave case, the second term in each equation represents
the correction, and the first term the value extracted using the
uncorrected physical amplitudes.  The corrections, at least for
\lamtxt and $\Xi$ decays, are model-independent at the $10-15\%$
level.  They are also very large, the ratio of corrected to uncorrected
values being $4.73(5.01)$ for
$\Lambda\rightarrow N\pi$ and $2.42(2.21)$
values for $\Xi\rightarrow \Lambda\pi$, for the models of Refs.~2,7 (8),
respectively.

	We conclude with a discussion of the decays $\oxmpo$ and
$\oxopm$, which are dominated by the PC p-wave process.  They are
expected to have very small baryon pole contributions$^{9-11}$,
and hence be dominated by the $K$ pole term.  Neglecting the
baryon pole term completely and recalling that the
$\pie$ $K$ pole contribution is $1/\sqth$ that for $\pit$
in leading order, we obtain
$$B(\oxmpt )=(1-\thetam /\sqth )B(\oxmpo )\ .\eqno(14)$$
The resulting change in the extracted \dith amplitude is only $+5.7\%$.  If we
use, instead, the results of Ref.~11 for the baryon pole and $K$ pole
contributions,
and the fact that the $10_F\rightarrow 8_F\times 8_F$ $\pie$ strong
coupling is $-\sqth$ times that for $\pit$, the correction term
in Eqn.~(14) is increased by a factor of $1.38$, leading to a net
change in the \dith amplitude of $+7.9\%$.  In either case the correction
is small.  This smallness results, first, from the small coefficient
in Eqn.~(14) and, second, from the fact that the nominal \dioh to
\dith ratio is much smaller in this case than for other hyperon decays.

It should be noted that, in making the estimates above, we have
ignored isospin-mixing due to electromagnetism (EM).  It is easy to
see that this is a rather good approximation.  First, the EM $\pit -\pie$
mixing is known to vanish at leading order in the chiral expansion$^{12}$,
and hence will be very small.  Second, using $U$-spin arguments, one may
derive$^{13}$, for $\lamo -\sigo$ mixing, the generalized Coleman-Glashow
relation
$$\delta m_{\lamo \sigo}^{EM}={1\over \sqth}\bigl[ \delta m_{\Sigma^o}
-\delta m_{\Sigma^+}-\delta m_n +\delta m_p\bigr] ^{EM}\ .\eqno(15)$$
If one then uses the estimates of Ref.~3 for the octet baryon EM
self-energies (based on the Cottingham formula), one finds
$\delta m_{\lamo\sigo}^{EM}=-0.09 $ MeV, which would alter $\theta_b$
by less than $8\%$.  Since such a shift is significantly smaller
than the $\simeq 20\%$ effects one might expect beyond leading
order in the quark masses, we neglect it.  We have, similarly,
neglected the effects of mixing between $\pit$ and $\pi_o$, where
$\pi_o$ is the SU(3) scalar member of the pseudoscalar nonet.  Again
one can see that this is likely to be a good approximation since,
for s-waves, the leading $\pi_o$ commutator terms vanish, while
for p-waves, using the pole model picture, the $K$ pole terms
remain small and the sum of the two distinct baryon pole terms vanishes
for each $\pi_o$ decay process as a result of the SU(3) scalar nature of the
$B^\prime B\pi_o$ strong couplings.  Combined with the fact that, using
quark model arguments, one expects the $\pit -\pi_o$ mixing angle to be
$\simeq 0.4\theta_m$, such contributions to the corrections should
be safely negligible.  (It should be noted that an analogous treatment
of particle mixing effects for the s-wave amplitudes was performed
previously by de la Torre$^{14}$.  The numerical values of the
corrections differ considerably from those obtained here.  The
origin of the difference is a very large EM contribution to $\lamo -\sigo$
mixing ($17$ times that obtained from Eqn.~(15)), which is $3$ times
as large as the mass mixing contribution and of opposite sign.  This
contribution is obtained using the quark model picture for the baryons,
together with the SU(3) limit of one photon exchange.  The resulting
EM contributions, however, do not satisfy the SU(3) relation Eqn.~(15).  The
situation is presumably similar to that of the pseudoscalar sector where
the analogous treatment, ignoring the class of photon loop graphs, fails
to satisfy the known chiral constraints$^{15}$ on the pseudoscalar
EM self-energies$^{16}$, e.~g., the vanishing of the $\pi_3$ EM self-energy
and $\pi_3-\pi_8$ EM mixing.)

In conclusion, we have demonstrated that corrections due to $\lamo -\sigo$
and $\pit -\pie$ mixing are required in order to extract the true
\dith transition amplitudes from experimental data on hyperon
non-leptonic decays.  The corrections are modest, though non-trivial, for
s-waves amplitudes, and extremely large, though somewhat model-dependent, for
p-wave amplitudes.  It is to the corrected values, and not those usually
extracted, that any attempts to model the \dith amplitudes must be
compared.
\vfill\eject
\
\vskip 1.5in
\noindent
{\bf ACKNOWLEDGEMENTS}
\vskip .15in
\noindent
I would like to thank John Donoghue for bringing the work of Ref.~(14)
to my attention.
The continuing support of the Natural Sciences and Engineering
Research Council of Canada, and the hospitality of the Department of
Physics and Mathematical Physics of the
University of Adelaide are also gratefully acknowledged.
\vfill\eject
\
\vskip 1in\noindent
{\bf REFERENCES}
\vskip .15in\noindent
1.  G. Karl, Phys. Lett. {\bf B328} (1994) 149
\newref
2.  J.F. Donoghue, E. Golowich and B. Holstein, Phys. Rep. {\bf 131}
(1986) 319
\newref
3.  J. Gasser and H. Leutwyler, Phys. Rep. {\bf 87} (1982) 77
\newref
4.  J.F. Donoghue, Ann. Rev. Nucl. Part. Sci. {\bf 39} (1989) 1
\newref
5.  Particle Data Group, Review of Particle Properties, Phys. Rev. {\bf D45}
(1992)
\newref
6.  A very clear exposition of the lowest order chiral analysis may be
found in J.F. Donoghue, E. Golowich and B. Holstein, ``Dynamics of the
Standard Model'', Cambridge University Press, New York, N.Y., 1992
\newref
7.  M. Gronau, Phys. Rev. Lett. {\bf 28} (1972) 188
\newref
8.  G. Nardulli, Il Nuov. Cim. {\bf 100A} (1988) 485
\newref
9.  J. Finjord, Phys. Lett. {\bf B76} (1978) 116
\newref
10. J. Finjord and M.K. Gaillard, Phys. Rev. {\bf D22} (1980) 778
\newref
11. M. Lusignoli and A. Pugliese, Phys. Lett. {\bf B132} (1983) 178
\newref
12. R. Dashen, Phys. Rev. {\bf 183} (1969) 1245
\newref
13. R.H. Dalitz and F. von Hippel, Phys. Lett. {\bf 10} (1964) 153
\newref
14. L. de la Torre, ``On Particle Mixing and Hypernuclear Decay'', Univ.
of Mass. PhD thesis, Sept. 1992
\newref
15. R. Dashen, Phys. Rev. {\bf 183} (1969) 1245
\newref
16. K. Maltman, G.J. Stephenson Jr. and T. Goldman, Nucl. Phys. {\bf A530}
(1991) 539
\vfill\eject
\
\vskip 1in
\noindent
Table 1.  Corrections to the s-wave amplitudes$^a$
\vskip .5in\noindent
\hrule
\vskip .10in\noindent
\settabs\+\qquad\qquad\qquad&Processspacespace\qquad\qquad&Experiment
\qquad\qquad&Correction&\cr
\+&Process&Experiment&Correction\cr
\vskip .1in\noindent
\hrule
\vskip .15in\noindent
\+&$\lppm$&\ 3.25&\ 0.048\cr
\+&$\lnpo$&-2.37&-0.028\cr
\+&$\spppo$&-3.27&-0.083\cr
\+&$\spnpp$&\ 0.13&\ \ \ 0\cr
\+&$\smnpm$&\ 4.27&\ \ \ 0\cr
\+&$\xlpm$&-4.51&-0.020\cr
\+&$\xlpo$&\ 3.43&\ 0.068\cr
\vskip .1in\noindent
\hrule
\vskip .15in\noindent
$^a$All entries in units of $10^{-7}$.  To obtain, e.~g. $A(\lonpt )$ one
adds the results of columns 2,3 for the process $\lnpo$.
\vfill\eject
\ \vskip .5in\noindent
Table 2.  Octet hyperon p-wave amplitudes$^a$
\vskip .5in\noindent
\hrule\vskip .1in\noindent
\settabs\+\qquad\qquad&Processspacespacespace&Model 1\qquad\qquad&Model 2
\qquad\qquad&Experiment&\cr
\+&Process&Model 1&Model 2&Experiment\cr
\vskip .1in\noindent\hrule\vskip .15in\noindent
\+&$\loppm$&\ 16.3&\ 17.9&\ 22.1\cr
\+&$\lonpt$&-11.4&-12.8&-15.8\cr
\+&$\spppt$&\ 20.3&\ 32.6&\ 26.6\cr
\+&$\spnpp$&\ 28.4&\ 45.8&\ 42.4\cr
\+&$\smnpm$&\ -0.8&\ -0.3&-1.44\cr
\+&$\xlopm$&\ 23.9&\ 13.3&\ 16.6\cr
\+&$\xlopt$&-17.0&\ -9.4&-12.3\cr
\vskip .15in\noindent
\+&$\soppm$&-26.6&-26.6&---\cr
\+&$\sonpt$&\ 20.4&\ 20.4&---\cr
\+&$\lonpe$&\ 44.5&\ 46.2&---\cr
\+&$\spppe$&-34.5&-20.0&---\cr
\+&$\xsopm$&-15.0&\ -3.6&---\cr
\+&$\xsopt$&-63.7&-43.5&---\cr
\+&$\xlope$&-20.9&\ -0.5&---\cr
\vskip .1in\noindent\hrule\vskip .15in\noindent
$^a$All entries in units of $10^{-7}$.  Models 1,2 are the models of
Refs.~2,7 and 8, respectively, and are described in the text.  Experimental
values refer, where listed, to the corresponding physical amplitude.
\vfill\eject
\
\vskip 1in\noindent
Table 3.  Corrections to the p-wave amplitudes$^a$
\vskip .5in\noindent\hrule\vskip .1in\noindent
\settabs\+\qquad\qquad&Processspacespace\qquad&Experiment
\qquad\qquad&Model 1\qquad\qquad&Model 2&\cr
\+&Process&Experiment&Model 1&Model 2\cr
\vskip .1in\noindent\hrule\vskip .15in\noindent
\+&$\lppm$&\ 22.1&-0.399&-0.399\cr
\+&$\lnpo$&-15.8&-0.362&-0.388\cr
\+&$\spppo$&\ 26.6&\ 0.518&\ 0.300\cr
\+&$\spnpp$&\ 42.4&\ \ \ 0&\ \ \ 0\cr
\+&$\smnpm$&-1.44&\ \ \ 0&\ \ \ 0\cr
\+&$\xlpm$&\ 16.6&-0.225&-0.053\cr
\+&$\xlpo$&-12.3&-0.642&-0.644\cr
\vskip .1in\noindent\hrule\vskip .15in\noindent
$^a$All entries in units of $10^{-7}$.  Models 1,2 are as described
in Table 2.  To obtain, e.~g. $B(\lonpt )$, one adds the entry of
column 2(or 3) to that of column 1 for the corresponding physical
process $\lnpo$.
\vfill\eject\bye